\documentclass[twocolumn,showpacs]{revtex4}
\usepackage{graphicx}
\usepackage{dcolumn}
\usepackage{amsmath}
\begin{document}
\title{Scaling properties of granular materials}
\author{Thorsten P\"oschel, Clara Salue\~na, and Thomas Schwager}
 \homepage{http://summa.physik.hu-berlin.de/~kies/}
 \affiliation{Humboldt-University Berlin -- Charit\'e,
   Monbijoustra{\ss}e 2, D-10117 Berlin, Germany} \date{\today}

\begin{abstract}
  Given an assembly of viscoelastic spheres with certain material
  properties, we raise the question how the macroscopic properties of
  the assembly will change if all lengths of the system, i.e. radii,
  container size etc., are scaled by a constant. The result leads to a
  method to scale down experiments to lab-size.
\end{abstract}
\pacs{45.70.-n, 45.70.Mg}
\maketitle

Assume the dynamics of a certain granular system $S$ is known. Will
the dynamics change if we rescale all sizes by a constant factor
$\alpha$, i.e., $R_i^\prime \equiv \alpha R_i$, but leaving the
material properties unchanged?\,\cite{prime}
If scaling affects the system properties, how do we have
to modify the material properties to assure that the system $S$ and
the scaled system $S^\prime$ behave identically?

It is frequently desired to investigate large scale phenomena in
granular systems experimentally, such as geophysical effects or
industrial applications. To this end one has to rescale all lengths of
the system to meet the restrictions of the laboratory size, i.e. big
boulders in the original system become centimeter sized particles in
the experiment. Of course, one wishes that the effects which occur in
the large system occur equivalently in the scaled system too. With the
assumption of viscoelastic particle deformation we will show that
naive scaling will modify the properties of a granular system such
that the original system and the scaled system might reveal quite
different dynamic properties. To guarantee equivalent dynamical
properties of the original and the scaled systems we have to modify
the material properties in accordance with the scaling factor and we
have to redefine the unit of time.

As another consequence of the scaling properties we claim that for
numerical simulations of granular material it is not sufficient to
provide relative data such as, e.g. to describe the container size in
units of the particle diameter. We will show an example where the
dynamics of a granular system changes significantly with system size,
although all relative sizes are kept constant.

In a simple approximation, a granular system may be described as an
assembly of spheres of radii $R_i$, $i=1,\dots,N$. If two particles of
radii $R_i$ and $R_j$ at positions $\vec{r}_i$ and $\vec{r}_j$ touch,
i.e., if $\xi_{ij}\equiv R_i+R_j-\left| \vec{r}_i - \vec{r}_j\right| >
0$, they feel an interaction force
\begin{equation}
  \label{eq:force}
  \vec{F}_{ij}=F^n_{ij}\vec{n}_{ij} + F^t_{ij}\vec{t}_{ij}\,,
\end{equation}
with the unit vector in normal direction $\vec{n}_{ij}\equiv
\left(\vec{r}_j-\vec{r}_i\right)/{\left|\vec{r}_j-\vec{r}_i\right|}$
and the respective unit vector in tangential direction $\vec{t}_{ij}$.
Eventually, external forces as, e.g., gravity, may also act on the
particles.

The normal force $F^n$ can be subdivided into elastic and dissipative
parts $F^n=F^n_{\rm el} + F^n_{\rm dis}$ \cite{indices}.
The elastic force for colliding spheres is
given by Hertz's law~\cite{Hertz}
\begin{equation}
  \label{eq:Hertz}
  F^n_{\rm el}=\frac{2Y}{3\left(1-\nu^2\right)}\sqrt{R^{\rm eff}} 
  \xi^{3/2}\equiv \kappa \xi^{3/2}\,,
\end{equation}
with $R^{\rm eff}=R_i R_j/ \left(R_i+R_j\right)$ and $Y$, $\nu$ being
the Young modulus and the Poisson ratio. Equation (\ref{eq:Hertz})
also defines the prefactor $\kappa$ which we will need below.

The formulation of the dissipative part of the force $F^n_{\rm dis}$
depends on the mechanism of damping. Here we will focus on
viscoelastic damping which is the most simple assumption for
dissipatively colliding bodies~\cite{brittle}.
It implies that the elastic part of the stress tensor
is a linear function of the deformation tensor and the dissipative
part of the stress tensor is a linear function of the deformation rate
tensor. It is valid if the characteristic velocity (the impact rate
$g$) is much smaller than the speed of sound $c$ in the material and
the viscous relaxation time $\tau_{\rm vis}$ is much smaller than the
duration of the collisions $\tau_c$~\cite{BSHP}.  The range of the
viscoelastic model is, hence, limited from both sides: the collisions
should not be too fast to assure $g\ll c$, $\tau_{\rm vis}\ll
\tau_{c}$, and not too slow to avoid influences of surface effects as
adhesion.  For viscoelastically colliding spheres the dissipative part
of the normal force reads~\cite{Kuwabara,BSHP,Morgato}
\begin{equation}
  F^n_{\rm dis} = A\frac{d\xi}{dt}\,\frac{d}{d\xi}F^n_{\rm el} = 
\frac32 A\kappa\sqrt{\xi}~\frac{d\xi}{dt}\,.
\label{eq:viscospheres}
\end{equation}
The dissipative material constant $A$ is a function of the viscous
constants $\eta_{1/2}$, the Young modulus $Y$ and the Poisson ratio
$\nu$ (for details see~\cite{BSHP}):
\begin{equation}
  \label{eq:A}
  A=\frac13\frac{\left(3\eta_2-\eta_1\right)^2}{3\eta_2+2\eta_1} 
\frac{\left(1-\nu\right)\left(1-2\nu\right)}{Y\nu^2}\,.
\end{equation}

Combining the forces (\ref{eq:Hertz}) and (\ref{eq:viscospheres}) one
obtains the equation of motion
\begin{equation}
  \label{eq:eom}
  \frac{d^2\xi}{dt^2} + \frac{\kappa}{m^{\rm eff}} \left(\xi^{3/2} + 
    \frac32 A\sqrt{\xi}~\frac{d\xi}{dt}\right) =0\,,
\end{equation}
with $m^{\rm eff}=m_im_j/(m_i+m_j)$ and with the initial conditions
$\left.\xi\right|_{t=0} = 0$ and $\left.\left.
    d\xi\right/dt\right|_{t=0} = g$.

In dimensionless variables, $\hat{\xi}\equiv \xi/\xi_0$, $\tau\equiv
t/ \tau_0$, Eq.~(\ref{eq:eom}) reads~\cite{DimAnalys}
\begin{equation}
  \label{eq:eomdimless}
  \frac{d^2\hat{\xi}}{d\tau^2} + \frac54\hat{\xi}^{3/2} + 
\frac32\left(\frac54\right)^{3/5} A\left(\frac{\kappa}
    {m^{\rm eff}}\right)^{2/5} g^{1/5}\sqrt{\hat{\xi}}~ 
  \frac{d\hat{\xi}}{d\tau} = 0\,,
\end{equation}
with $\left.\hat{\xi}\right|_{\tau=0} = 0$ and
$\left.\left.d\hat{\xi}\right/d\tau\right|_{\tau=0} = 1$. The
characteristic length $\xi_0$ is the maximal compression for the
equivalent undamped (elastic) problem which can be found by equating
the kinetic energy of the impact $m^{\rm eff}g^2/2 $ and the elastic
energy at the instant of maximal compression $2 \kappa\xi_0^{5/2}/5$.
As characteristic time $\tau_0$ we define the time in which the
distance between the particles changes by the characteristic length
just before the collision starts~\cite{colltime}
\begin{equation}
  \label{eq:charlengthtime}
\xi_0\equiv\left(\frac54\frac{m^{\rm eff}}{\kappa}\right)^{2/5} g^{4/5}\,
,~~~~~~~~  \tau_0\equiv \xi_0/g\,.
\end{equation}
The only term in Eq.~(\ref{eq:eomdimless}) which depends explicitly on
the system size and on material properties is the prefactor in front
of the third term. If the scaling procedure affects the value of this
term it will change the dynamics of the system. To ensure identical
behavior of the scaled system, however, besides the identity of this
prefactor, further requirements have to be met which will be discussed
below.

Expanding our abbreviations we obtain
\begin{eqnarray}
&&  A\left(\frac{\kappa}{m^{\rm eff}}\right)^{2/5} g^{1/5}\nonumber \\
&&=\left(2\pi\right)^{-2/5} 
A  F(R_i,R_j)\, g^{1/5} \left(\frac{Y}
{\rho\left(1-\nu^2\right)}\right)^{2/5}\,,
  \label{eq:prefactor1}
\end{eqnarray}
with $\rho$ being the material density. The function $F(R_i,R_j)$ 
 collects all terms containing $R_i$ and $R_j$:
 \begin{equation}
   \label{eq:F}
   F(R_i,R_j) \equiv \frac{\left[R_iR_j/\left(R_i+R_j\right)\right]^{1/5}}
{\left[R_i^3R_j^3/
      \left(R_i^3+R_j^3\right)\right]^{2/5}}
\end{equation}
Scaling the radii by $\alpha$, the function $F$ scales
$F(R_i^\prime,R_j^\prime) = F(\alpha R_i, \alpha R_j) = \alpha^{-1}
F(R_i,R_j)$. Obviously, simple scaling of the system in general
affects the prefactor Eq.~(\ref{eq:prefactor1}) already via the
scaling properties of $F\left(R_i,R_j\right)$, hence, in general the
original system and the system where the lengths have been scaled by a
factor $\alpha$ differ in their dynamic properties.  More explicitly,
one can show that naively scaling the system by a factor $\alpha < 1$
will lead to a comparatively more damped dynamics.
 
To provide equivalent dynamical properties of the systems, therefore,
we have to modify the material properties in a way to assure that the
equations of motion of both systems are equivalent, which in turn
assures that the prefactors (\ref{eq:prefactor1}) of the original
system and the scaled system are identical.

One of the few things which cannot be modified in experiments with
reasonable effort is the constant of gravity $G$.  That implies that
going from $S$ to $S^\prime$ not only $G$ but all other accelerations
must remain unaffected too
\begin{equation}
  \label{eq:accel}
  \left(\frac{d^{\,2}x}{dt^2}\right)^\prime = \frac{d^{\,2} (\alpha x)}
{d\left(t^\prime\right)^2} = \frac{d^{\,2}x}{dt^2}\,,
\end{equation}
yielding $t^\prime = \sqrt{\alpha}\,t$. Hence, scaling all lengths
$x^\prime = \alpha x$ implies that times scale as $t^\prime=
\sqrt{\alpha}t $ if we require that the gravity constant stays
unaffected. Thus, the clock in the scaled system $S^\prime$ must run
by a factor $\sqrt{\alpha}$ faster (or slower) than the clock in the
original system. In other words, if in the original system we observe
a phenomenon at time $t=100$\,s, we will find the same effect in the
scaled system at time $t^\prime=\sqrt{\alpha}\, 100$\,s. Scaling of
time is a direct consequence of scaling the lengths if the constant of
gravity has the same value in both systems.

In the scaled system the equation of motion of a particle contact reads
\begin{equation}
   \frac{d^2\xi^\prime}{dt^{\prime2}} +
\frac{\kappa^{\prime}}{\left(m^{\rm eff}\right)^\prime}
\left(\xi^\prime\right)^{3/2} +
    \frac32 A^{^\prime}
\frac{\kappa^{\prime}}{\left(m^{\rm eff}\right)^\prime}
   \sqrt{\xi^{\prime}}~\frac{d\xi^{\prime}}{dt^{\prime}}=0\,.
\label{eq:scaledeom}
\end{equation}
If we apply our scaling relations which were introduced above, i.e. 
\begin{equation}
   \xi^{\prime}=\alpha\xi\,,~~~~~~~
   \frac{d\xi^{\prime}}{dt^{\prime}}=\sqrt{\alpha}\frac{d\xi}{dt}\,,~~~~~~~
   \frac{d^2\xi^{\prime}}{dt^{\prime 2}}=\frac{d^2\xi}{dt^2}
   \label{eq:velscale}
\end{equation}
\noindent we obtain 
\begin{equation}
  \frac{d^2\xi}{dt^2} + \alpha^{3/2}
\frac{\kappa^{\prime}}{\left(m^{\rm eff}\right)^\prime}
\xi^{3/2} + \frac32 \alpha A^{\prime}
\frac{\kappa^{\prime}}{\left(m^{\rm eff}\right)^\prime}
\sqrt{\xi}~\frac{d\xi}{dt}=0\,.
\end{equation}
Comparing with Eq.~(\ref{eq:eom}) we find the conditions to assure
identity of the equations of motion:
\begin{equation}
\frac{\kappa^{\prime}}{\left(m^{\rm eff}\right)^\prime}=\alpha^{-3/2}
\frac{\kappa}{m^{\rm eff}}\,,~~~~~   A^{\prime}=\sqrt{\alpha}\,A\,.
\label{eq:betterscale}
\end{equation}
Using the definitions of $\kappa$ (Eq.~(\ref{eq:Hertz})) and $m^{\rm
  eff}$ yields finally
\begin{equation}
  \label{eq:betterscale1}
\left(\frac{Y}{\rho \left(1-\nu^2\right)}\right)^\prime = 
\alpha \frac{Y}{\rho \left(1-\nu^2\right)} \,,~~~~~   
A^{\prime}=\sqrt{\alpha}\,A\,. 
\end{equation}
If we choose material constants which obey Eqs.
(\ref{eq:betterscale1}) we will obtain the original equation of motion
after scaling the system back to its original size, i.e., both systems
are equivalent.
 
When we incorporate the tangential force $F^t$ into the analysis, of
course, we have to require that this force scales in the same way as
any other force, namely
\begin{equation}
  \frac{\left(F^t\right)^\prime}{\left(m^{\rm eff}\right)^\prime}
=\frac{F^t}{m^{\rm eff}}\,,
\end{equation}
given that accelerations are invariant under scaling.  This
requirement has to be met by appropriately scaling the material
constants, particularly the friction constant, resulting in an
additional scaling equation. Its form depends on the underlying
friction model, i.e. on the functional dependence of the tangential
force on the geometry and the material properties as well as on the
compression and the relative velocity. For instance, if we assume the
most simple tangential force law
\begin{equation}
  F^t=\mu F^n\,,
\end{equation}
with $\mu$ being the friction coefficient we can conclude that the
friction coefficient has to be invariant with respect to scaling
\begin{equation}
  \mu^\prime = \mu\,.
\label{eq:mu}
\end{equation}

In literature, one can find a variety of different models which might
be more realistic for the description of the dynamics of a granular
material (for an overview see, e.g., \cite{wolf}). Applying the same
procedure for these laws will result in different scaling relations
and, hence, in different conditions of the type Eq.~(\ref{eq:mu}). A
thorough discussion of the scaling of more realistic tangential
friction laws will be published elsewhere.

\begin{table}
\caption{Material parameters used in the simulations}
\label{tab:material}
\begin{ruledtabular}
  \begin{tabular*}{\hsize}{ l l l l}
   samples           & ~~original & ~~scaled I & ~~scaled II  \\ \hline
   plane length      & ~~10 m     & ~~62.5 cm  & ~~160 m \\
   particle diameter & ~~10 cm    & ~~6.25 mm  & ~~160 cm     \\
   $\rho$            & ~~2 g/cm$^3$ & ~~2 g/cm$^3$  & ~~2 g/cm$^3$ \\
   $\frac{\displaystyle Y} {\displaystyle \left(1-\nu^2\right)}$ & 
~~8 GPa & ~~0.5 GPa  & ~~128 GPa \\[0.3cm]
   $A$               & ~~2$\cdot 10^{-5}$ s &~~ $5\cdot 10^{-6}$ s  & 
~~$8\cdot 10^{-5}$ s\\
   $\mu$             & ~~0.1      & ~~0.1      & ~~0.1  
  \end{tabular*}
\end{ruledtabular}
\end{table}

\begin{figure}[htb]
\includegraphics[width=8cm]{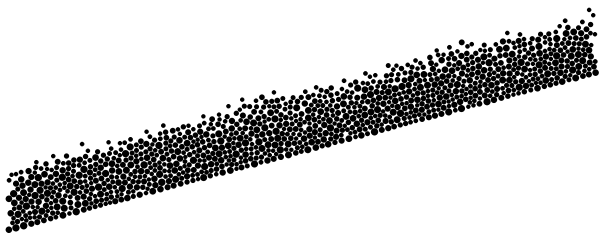}
\vspace*{0.3cm}
\caption{Snapshot of the granular flow down an inclined plane.}
  \label{fig:snapshot}
\end{figure}

To demonstrate the derived scaling laws, we present results of a 2D
Molecular Dynamics simulations of stationary flow of 1000 particles
down an inclined plane. In the ``original system'' the length of the
plane is 10\,m (periodic b.c.), its slope is 15 degrees, and the
particle diameters have been randomly chosen with an average of
10\,cm. The material parameters are given in Table~\ref{tab:material}.
Particles are subjected to normal viscoelastic forces as well as
tangential forces (of the simple type discussed above) which, although
not very realistic, is enough for the purpose of showing the scaling
properties.  Figure~\ref{fig:snapshot} shows a snapshot of the flow.

\begin{figure}[htb]
\includegraphics[width=5cm,angle=270]{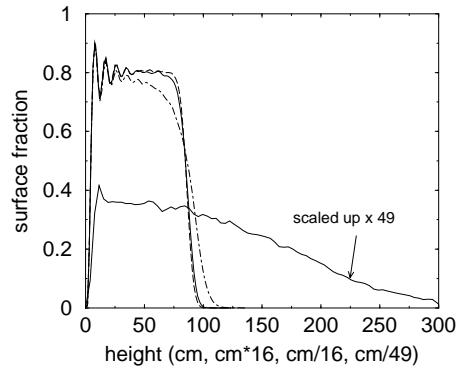}
\caption{Density profile of the flow down an inclined plane. Thick 
  solid line: original system together with scaled systems I and II
  ($\alpha_I=1/16$, $\alpha_{II}=16$, correct scaling, parameters see
  Table \ref{tab:material}). The curves collapse precisely to one
  single line. Dashed line: scaled system II, $\alpha=1/16$ (incorrect
  scaling), dot-dashed line: scaled system, $\alpha=16$ (incorrect
  scaling). Thin solid line: system incorrectly scaled by $\alpha=49$.
  For the scaled systems, the horizontal axis has been multiplied by
  16 (I), 1/16 (II) and 1/49.}
  \label{fig:avgprof}
\end{figure}

The solid curve in Fig. \ref{fig:avgprof} represents the density
profile perpendicular to the inclined plane of the original system and
of the correctly scaled systems I and II, taking into account the
necessary modification of the material parameters according to
Eqs.~(\ref{eq:betterscale1},\ref{eq:mu}). The modified material
parameters are summarized in Table \ref{tab:material}. It is not
surprising that all three lines collapse perfectly, since in scaled
variables the equations of motion of the original system and the
correctly scaled system are identical. Therefore, particle
trajectories and, hence, the density plot must be identical.

Figure \ref{fig:avgprof} also shows two curves for the stationary
density profile of the systems scaled by $\alpha=1/16$ (system I) and
$\alpha=16$ (system II) where only the lengths have been scaled but
erroneously the material parameters have been kept constant (incorrect
scaling). These curves deviate from the profile of the original
systems since the original and the scaled systems obey different
equations of motion.  As anticipated, the scaling by $\alpha=16$ only
applied to lengths leads to a more elastic dynamics, i.e.,
comparatively more dilute flow (dot-dashed curve in
Fig.~\ref{fig:avgprof}).  This effect is amplified when larger scaling
factors are applied ($\alpha=49$), where in twice the simulation time,
no steady state is found while the density profile keeps relaxing
towards more dilute states. The instantaneous density profile is shown
as a thin-solid curve in Fig.~\ref{fig:avgprof}, as an indication that
incorrect scaling can ultimately lead to completely deviated results.

The simulations show that the simple length scaling of a granular
system by a constant factor $\alpha$ changes the dynamical properties
significantly, if the material parameters are kept constant. In order
to obtain identical results for a scaled system one has also to modify
the material constants and redefine the unit of time. These necessary
scaling relations are summarized in Table \ref{tab:scale}.
\begin{table}
\caption{The necessary scaling relations when transiting from a granular 
system ($S$) to a scaled system ($S^{\prime}$) having identical 
dynamic properties}
\label{tab:scale}
\begin{ruledtabular}
  \begin{tabular}{l|l@{~~~~~}l}
   &~~ original system & scaled system\\ \hline
   all lengths &~~ $x$ & $\alpha\,x$\\[0.1cm]
   time &~~ $t$ & $\sqrt{\alpha}\,t$ \\[0.1cm]
   elastic constant &~~ $\frac{\displaystyle Y}{\displaystyle \rho 
\left(1-\nu^2\right)}$ & 
   $\alpha\, \frac{\displaystyle Y}{\displaystyle \rho 
\left(1-\nu^2\right)}$ \\[0.3cm]
   dissipative constant~~ &~~ $A$ & $\sqrt{\alpha}A$
  \end{tabular}
\end{ruledtabular}
\end{table}

The knowledge about these scaling relations offers the possibility to
scale down real world systems, e.g.  geophysical or industrial
granular systems, to sizes where laboratory experiments can be
performed. If one scales down such a granular system one has to
replace the original material by a material which meets the scaling
requirements discussed in the text.

We want to give an example: Assume in the original system one deals
with steel spheres ($Y=20.6\cdot 10^{10}$ Nm$^{-2}$, $\nu=0.29$ and
$\rho=7,700\,$kg m$^{-3}$) of average radius $\bar{R}=10$\,cm and
system size of $L=10$\,m. The property whose scaling behavior is known
is $Y/(\rho(1-\nu^2))=2.92\cdot 10^7$ m$^2$s$^{-2}$. One wishes to
know (to measure) a certain value at time $t=100$\,s. In the lab we
perform the experiment with an equivalent system of size $L^\prime =
1$ m, i.e. we scale the system by the factor $\alpha=0.1$, including
all radii.  From the scaling relations we see that we have to find a
material whose scaled property is
$Y^{\prime}/(\rho^{\prime}(1-\nu^{\prime 2}))\approx0.3\cdot 10^7$
m$^2$s$^{-2}$. From tables~\cite{tables} we find that we can use
plexiglass ($Y=0.32\cdot 10^{10}$ Nm$^{-2}$, $\nu=0.35$ and
$\rho=1,200$ kg m$^{-3}$) in order to obtain this value. Therefore, we
have to perform the experiment with plexiglass spheres and have to
measure the value of interest at time $t^\prime = 31.6$\,s.

One can imagine that not for all scaling factors $\alpha$ one will
find a proper material, however, nowadays it is possible to
manufacture materials which can meet demanding requirements, such as
high softness along with a custom-designed damping constant.

The scaling properties have also consequences for Molecular Dynamics
simulations of granular systems: namely, it is not sufficient in
simulations to provide {\em relative} parameters such as the container
size in units of the particle radius. As demonstrated, the result of a
simulation (and, of course, also of a real world experiment) depends
on {\em absolute} values.  \medskip

\begin{acknowledgments}
  This work has been supported by Deutsche Forschungsgemeinschaft via
  grants PO 472/6-1, PO 472/7-1.
\end{acknowledgments}

\begin{thebibliography}{1}
\bibitem{prime} Here and in the following we mark all variables which
  describe the scaled system $S^{\prime}$ with a prime.
\bibitem{indices} For simplicity of notation we drop the indices $ij$
  of the variables which refer to pairs of particles.
\bibitem{colltime} Note that up to a numerical prefactor the timescale
  $\tau _0$ is equal to the duration of the undamped collision \cite
  {Hertz}, which would be an alternative (and equivalent) choice of
  the timescale.
\end{thebibliography}

\end{document}